\documentstyle[12pt]{article}
\font\twlmsb =msbm10 scaled \magstep1
\font\egtmsb =msbm8
\font\sevmsb =msbm7
\newfam\msbfam
\textfont\msbfam\twlmsb
\scriptfont\msbfam\egtmsb
\scriptscriptfont\msbfam\sevmsb
\def\Bbb{\protect\pBbb}
\def\pBbb{\relax\ifmmode\expandafter\Bb\else\typeout{You cann't use
Bbb in text mode}\fi}
\def\Bb #1{{\fam\msbfam\relax#1}}
\def\thebibliography#1{\bigskip\section*{\centering
References\\}\bigskip\list
{\arabic{enumi}.}{\settowidth\labelwidth{#1}\leftmargin\labelwidth
\advance\leftmargin\labelsep\usecounter{enumi}}
\def\newblock{\hskip .11em plus .33em minus .07em}
\sloppy\clubpenalty4000\widowpenalty4000 \sfcode`\.=1000\relax}

\def\op#1{\mathop{\fam0 #1}\limits}

\newcommand{\ben}{\begin{eqnarray}}
\newcommand{\een}{\end{eqnarray}}
\newcommand{\be}{\begin{eqnarray*}}
\newcommand{\ee}{\end{eqnarray*}}
\newcommand{\bea}{\begin{eqalph}}
\newcommand{\eea}{\end{eqalph}}

\newcommand{\al}{\alpha}

\newcommand{\dl}{\delta}
\newcommand{\la}{\lambda}
\newcommand{\La}{\Lambda}
\newcommand{\om}{\omega}

\newcommand{\g}{\gamma}

\newcommand{\e}{\epsilon}
\newcommand{\th}{\theta}

\newcommand{\f}{\phi}

\newcommand{\wt}{\widetilde}
\newcommand{\wh}{\widehat}
\newcommand{\ol}{\overline}
\newcommand{\id}{{\rm id\,}}
\newcommand{\dr}{\partial}

\newcounter{eqalph}
\newcounter{equationa}

\newenvironment{eqalph}{\stepcounter{equation}
\setcounter{equationa}{\value{equation}}
\setcounter{equation}{0}

\begin{eqnarray}}{\end{eqnarray}
\setcounter{equation}{\value{equationa}}}

\begin{document}
\hbox{}
\vskip15mm

\centerline{\large\bf True Functional Integrals}

\centerline{\large\bf in Algebraic Quantum Field Theory}
\medskip

\centerline{\bf G.Sardanashvily}
\medskip

\centerline{Department of Theoretical Physics, Physics Faculty}

\centerline{Moscow State University, 117234, Moscow, Russia}

\centerline{E-mail: sard@grav.phys.msu.su}
\vskip1cm

The familiar generating functional
\begin{equation}
N^{-1}\exp [i\int L(\phi)]\op\prod_x [d\phi(x)] \label{8.1}
\end{equation}
in QFT is introduced usually as formal extrapolation of measures
in finite-dimensional spaces. However, this fails to be a true measure since
the Lebesgue measure
\[
\op\prod_x [d\phi(x)]
\]
in infinite-dimensional linear spaces is not defined in general. Therefore,
the properties of the generating functional (\ref{8.1}) remain elusive
what becomes essential when describing non-perturbative phenomena.
The problem lies in constructing representations of topological
$^*$-algebras of quantum fields which are not normed \cite{bor,hor}.

These difficulties may be overcome if we restrict our
consideration only to chronological forms on quantum field algebras. In
this case, since chronological forms of boson fields are symmetric, the
algebra of quantum fields can be replaced with the commutative tenzor algebra
of the corresponding infinite-dimensional nuclear space $\Phi$. This is the
enveloping algebra of the abelian Lie group of translations in $\Phi$. The
generating functions of unitary representations of this group
\cite{gel} play the role
of Euclidean generating functionals in algebraic QFT. They are the Fourier
transforms of measures in the dual $\Phi'$ to the space $\Phi$
\cite{cam,S91,S91a}. To describe realistic field models, one can construct
these measures in terms of multimomentum
canonical variables when momenta correspond to derivatives of fields with
respect to all world coordinates\cite{sard9,sard10,int}. In the present work,
we spell out the algebraic aspects of the above-mentioned approach. As a test
case, scalar field is examined. By analogy with
the case of boson fields, the corresponding anticommutative
algebra of fermion fields is defined to be the algebra of functions
taking their values into an infinite Grassman algebra \cite{S90}.

\section{Quantum field algebras}

In accordance with the algebraic approach, a quantum field system can
be characterized by a topological $^*$-algebra $A$ (with the unit
element $1_A$) and a continuous positive form (state) $f$ on $A$, that is,
\[
f(aa^*)\geq 0, \qquad f(1_A)=1, \qquad a\in A.
\]
With reference to the field-particle dualism, realistic models are
described by tensor algebras as a rule, and states are treated as
vacuum expectation form.

Let $\Phi$ be a real linear locally convex topological space endowed
with an involution operation
\[
*:\f\to\f^*,\qquad \f,\f'\in\Phi.
\]
The linear space
\[
A_\Phi=\op\oplus^\infty_0\Phi_n, \qquad \Phi_0=\Bbb R, \qquad
\Phi_{n>0}=\op\otimes^n\Phi
\]
is a $\Bbb Z$-graded $^*$-algebra with respect to the natural tensor product
and the involution operation
\[
(\f_1\cdots\f_n)^*=(\f_n)^*\cdots(\f_1)^*,\qquad \f^i\in\Phi.
\]
The algebra $A_\Phi$ can be provided with the direct sum topology when
the sequence
\[
\{a^i=(\f^i_0,...,\f^i_n,...), \, \f^i_n\in\Phi\}
\]
converges to 0 if, whenever $n$, the sequence $\{\f^i_n\}$ converges to 0 with
respect to the topology of $\Phi_n$  and there exists a number N independent
of $i$ such that $\f^i_n=0$ for all $i$ and $n>N$. This topology bring
$A_\Phi$ into a topological $^*$-algebra.

A state $f$ on the tensor algebra $A_\Phi$ is given by the collection
of continuous forms $\{f_n\}$ on spaces $\Phi_n$.

We further assume that $\Phi$ is a nuclear space.

In the axiomatic quantum field theory of real scalar fields,
the quantum field algebra is the Borchers algebra $A_\Phi$ where
$\Phi=\Bbb RS_4$. By $\Bbb RS_m$ is meant the real subspace of the nuclear
Schwartz space $S(\Bbb R^m)$
of complex $C^\infty$ functions $\f(x)$ on ${\Bbb R}^m$ such that
\[
\|\f\|_{k,l}=\op{\max\ \ \sup}_{|\al|\leq l,\, x\in{\Bbb R}^m}(1+|x|)^k
\frac{\dr^{|\al|}}{(\dr x^1)^{\al_1}\ldots(\dr x^m)^{\al_m}}\f (x),
\]
\[
|\al|=\al_1+\cdots+\alpha_m,
\]
is finite for any collection $(\al_1,\ldots,\al_m)$ and all
$l,k\in{\Bbb Z}^{\geq 0}$. The space ${\Bbb R}S_m$ is endowed with the
set of seminorms $\|\f\|_{k,l}$ and the associated topology. It is
reflexive. The dual to $S^(\Bbb R^n)$ is the space of temperate
distributions (or simply distributions) $S'(\Bbb R^{4n})$.

The following properties of $S(\Bbb R^n)$ motivates us to choose it in
order to construct quantum field algebras. The translation operator
\[
x\to x+a,\qquad x\in\Bbb R^n,
\]
in $S(\Bbb R^n)$ has the complete set of generalized eigen vectors
$\exp (ipx)\in S'(\Bbb R^n)$. Moreover, $S(\Bbb R^n)$ and $S'(\Bbb
R^n)$ are isomorphic to their Fourier transforms $S(\Bbb R_n)$ and
$S'(\Bbb R_n)$ respectively:
\be
&& \f^F(p)=\int \f(x)\exp(ipx)d^nx,\\
&& \f (x)=\int \f^F(p)\exp(-ipx)d_np
\ee
where $\Bbb R_n =(\Bbb R^n)'$ and $d_n =(2\pi)^{-2n}d^n$.

Remark that the subset $\op\otimes^kS(\Bbb R^n)$ is dense in
$S(\Bbb R^{kn})$ and every continuous form on it is uniquely
extended to a continuous form on $S(\Bbb R^{kn})$. Every bilinear
functional $M(\f_1,\f_2)$ which is separately continuous with respect to
$\f_1\in S(\Bbb r^n)$ and $\f_2\in S(\Bbb R^m)$ is expressed uniquely
into the form
\[
M(\f_1,\f_2)= \int F(x,y)\f_1(x)\f_2(y)d^nxd^my,
\qquad F\in S'(\Bbb R^{n+m}).
\]
When provided with the nondegenerate separately continuous Hermitian form
\begin{equation}
\langle \f_1|\f_2\rangle =\int \f_1(x)\ol\f_2(x)d^nx, \label{S2}
\end{equation}
the nuclear space $S(\Bbb R^n)$ is the rigged Hilbert space.

As a consequence, every state $f=\{f_n\}$ on the Borchers algebra
$A_{\Bbb RS_4}$ is represented by a
family of temperate distributions $W_n\in S'(\Bbb R^{4n})$:
\[
f(\f_1\cdots\f_n)=
\int W_n(x_1,\ldots,x_n)\f_1(x_1)\cdots\f_n(x_n)d^4x_1\cdots d^4x_n.
\]
In particular, if $f$ obeys the
Wightman axioms, $W_n$ are the familiar $n$-point Wightman functions.

Let us consider the states on the Borchers algebra $A_{\Bbb RS_4}$
which describe free massive scalar fields.

Set the following complex positive bilinear form on $A_{\Bbb RS_4}$:
\begin{equation}
(\f|\f')=\frac2{i}\int d^4xd^4y\f(x)D^-(x-y)\f(y)=
\int\frac{d^3p}{\om}\f^F(\om,\ol p){\ol \f'}^F(\om, \ol p),
\label{S3}
\end{equation}
\[
D^-(x) = i(2\pi)^{-3}\int d^4p\exp(-ipx)\th(p_0)\dl (p^2-m^2),
\]
where $\th$ is the step function, $p^2$ the Minkowski scalar and
\[
\om=(\ol p^2+m^2)^{1/2}.
\]
Recall that, since the functions $\f$ are real, we have
\[
\f^F(p)=\ol\f^F(-p).
\]

The form (\ref{S3}) is degenerate, for $D^-$ satisfies the massive surface
equation
\[
(\Box +m^2)D^-(x)=0.
\]
Let us consider the quotient
\[
\g: \Bbb RS_4\to\Bbb RS_4/I
\]
of $\Bbb RS_4$ by the kernal
\[
I:= \{\f\in \Bbb RS_4:\, (\f|\f)=0\}
\]
of the form (\ref{S3}).
The transformation $\g$ sends every element $\f\in \Bbb RS_4$ whose
Fourier transform is $\f^F(p_0,\ol p)\in S(\Bbb R_4)$ to the pair of
functions $(\f^F(\om,\ol p),\f^F(-\om,\ol p))$. Set the bilinear form
\begin{equation}
(\g\f|\g\f')_L= {\rm Re}(\f|\f') =\frac12\int\frac{d^3\ol p}{\om}
[\f(\om,\ol p)\f'(-\om,-\ol p)+\f(-\om,\ol p)\f'(\om,-\ol p)].
\label{S1}
\end{equation}
The space $\Bbb RS_4/I$ is the direct sum of the subspaces
\[
L^\pm= \{\f^F_\pm(\om,\ol p) = \pm\f_\pm(-\om,\ol p)\}
\]
which are orthonomal to each other with respect to the form (\ref{S1}).

There are exist continuous isometric mappings
\be
&&\g_+: \f^F_+(\om,\ol p)\to q(\ol p)=\om^{-1/2}\f^F_+(\om,\ol p),\\
&&\g_-: \f^F_-(\om,\ol p)\to q(\ol p)=-i\om^{-1/2}\f^F_-(\om,\ol p)
\ee
of spaces $L^+$ and $L^-$ to the space $\Bbb RS_3$ endowed with the
separately continuous positive nondegenerate Hermitiam form
\[
\langle q|q'\rangle =\int d^3pq^F(\ol p){\ol q'}^F(\ol p).
\]
Note that the images of $\g_+(L^+)$ and $\g_-(L^-)$ in $\Bbb RS_3$ fail
to be orthonomal. The superpositions of $\g$ and $\g_\mp$ yield the
mappings
\be
&& \tau_+(\f) =\g_+(\g\f)_+ =\frac{1}{2\om^{1/2}}\int[\f^F(\om,\ol p) +
\f^F(-\om,\ol p)]\exp(-i\ol p\ol x)d^3p,\\
&& \tau_-(\f) =\g_+(\g\f)_- =\frac{1}{2i\om^{1/2}}\int[\f^F(\om,\ol p) -
\f^F(-\om,\ol p)]\exp(-i\ol p\ol x)d^3p,
\ee
of the space $\Bbb RS_4$ to $\Bbb RS_3$.

Now, let us consider the algebra CCR
\[
K_{\Bbb RS_3} = \{\pi(q), s(q), \, q\in \Bbb RS_3\}
\]
and set the morphism
\[
\Bbb RS_4 \ni \f\to a(\f)= s(\tau_+(\f))-\pi(\tau_-(\f)).
\]
One can think of the algebra $K_{\Bbb RS_3}$ as beeing the algebra of
simultaneous canonical commutation relations of fields $\f\in\Bbb RS_4$.
Every representation of this algebra CCR \cite{gel} determines the
positive form
\begin{equation}
f(\f_1...\f_n)=\int[s(\tau_+(\f_1))-\pi(\tau_-(\f_1))]\cdots
[s(\tau_+(\f_n))-\pi(\tau_-(\f_n))]d\mu \label{S6}
\end{equation}
on the Borchers algebra $A_{\Bbb RS_4}$. The corresponding
distributions $W_n$ obey the massive surface equation, and the following
commutation relations hold:
\[
W_2(x,y)-W_2(y,x)=-iD(x-y)
\]
where
\[
D(x)=i(2\pi)^{-3}\int d^4p[\th(p_0)-\th(-p_0)]\dl(p^2-m^2)\exp(-ipx)
\]
is the Pauli-Iordan function. It follows that the form (\ref{S6}) describes
massive scalar fields. In case of the Fock representation $\{\pi(q),
s(q)\}$ of the algebra CCR $K_{\Bbb RS_3}$, the form (\ref{S6}) obeys the
Wick relations
\[
f_n(\f_1\cdots\f_n)=\sum f_2(\f_{i_1}\f_{i_2})\cdots
f_2(\f_{i_{n-1}}\f_{i_n})
\]
where the sum is over all pairs $i_1<i_2,...,i_{n-1}<i_n$ and $f_2$
is the Wightman function
\[
W_2(x,y) =\frac{1}{i}D^-(x-y).
\]
In this case, the form (\ref{S6}) describes free massive scalar fields.

\section{Chronological forms}

In QFT, particles created at some moment and
destructed at another moment are described by the chronological forms $f^c$
on the Borchers algebra $A_{\Bbb RS_4}$. They are given by the expressions
\begin{equation}
W^c_n(x_1,\ldots,x_n)=
\op\sum_{(i_1\ldots i_n)}\th(x^0_{i_1}-x^0_{i_2})
\cdots\th(x^0_{i_{n-1}}-x^0_{i_n})W_n(x_1,\ldots,x_n) \label{030}
\end{equation}
where $(i_1\ldots i_n)$ is a rearrangement of numbers $1,\ldots,n$.
However, the forms (\ref{030}) fail to be distributions in general and
so, $f^c$ are
not continuous forms on $A_{{\bf R}S_4}$. For instance, $W^c_1\in
S'(\Bbb R)$ iff $W_1\in S'(\Bbb R_\infty)$ where by $\Bbb R_\infty$ is
meant the compactification of $\Bbb R$ by identification of the points
$\{\infty\}$ and $\{-\infty\}$. Moreover, the chronological forms $f^c$
are not positive. Thus, they are not states on the Borchers algebra.
At the same time, they issue from the Wick rotation of the
Euclidean states on $A_{{\bf R}S_4}$ (see Appendix) which describe particles
in the interaction zone.

Since chronological forms (\ref{030}) are symmetric,
the Euclidean states on the Borchers algebra $A_{\Bbb RS_4}$
can be introduced as states on the corresponding commutative tensor
algebra $B_{\Bbb RS_4}$. Note that they differ
>from the Swinger functions associated with the Wightman functions in
the framework of the constructive Euclidean quantum field theory
\cite{sim}.

Let $B_\Phi$ be the complexified quotient of $A_\Phi$ by the ideal whose
generating set is
\[
J :=\{\f\f'-\f'\f, \,\, \f,\f'\in\Phi\}.
\]
This algebra is exactly the enveloping algebra of the Lie algebra
associated with the commutative Lie group $G_\Phi$ of translations in
$\Phi$. We therefore can construct a state on the algebra
$B_\Phi$ as a vector form
of its cyclic representation induced by a strong-continuous unitary
cyclic irreducible representation of $G_\Phi$. Such a representation is
characterized by a positive-type continuous generating function $Z$ on $\Phi$,
that is,
\[
Z(\f_i-\f_j)\al^i\ol\al^j\geq 0, \qquad Z(0)=1,
\]
for all collections of $\f_1,\ldots,\f_n$ and
complex numbers $\al^1,\ldots,\al^n$. If the function
\[
\al\to Z(\al\f)
\]
on $\Bbb R$ is analytic at 0 for each $\f\in\Phi$, the state $F$ on $B_\Phi$
is given by the expressions
\[
F_n(\f_1\cdots\f_n)=i^{-n}\frac{\dr}{\dr\al^1}
\cdots\frac{\dr}{\dr\al^n}Z(\al^i\f_i)|_{\al^i=0}.
\]

In virtue of the well-known theorem \cite{gel}, any
function $Z$ of the above-mentioned type is the Fourier
transform of a positive bounded measure $\mu$
in the dual $\Phi'$ to $\Phi$:
\begin{equation}
Z(\phi)=\op\int_{\Phi'}\exp[i\langle w,\f\rangle]d\mu(w) \label{031}
\end{equation}
where $\langle,\rangle$ denotes the contraction between $\Phi'$ and
$\Phi$. The corresponding representation of $G_\Phi$ is given by operators
\[
\pi_\mu(g(\f)): u(w)\to\exp[i\langle w,\f\rangle]u(w)
\]
in the space $L^2(\Phi',\mu)$ of quadraticly $\mu$-integrable
functions $u(w)$ on $\Phi'$. We have
\[
F_n(\f_1\cdots\f_n)=\int\langle w,\f_1\rangle\cdots\langle w,\f_n
\rangle d\mu(w).
\]

It should be noted that the representations $\pi_\mu$ and $\pi_{\mu'}$
are not equivalent iff the measures $\mu$ and $\mu'$ are not equivalent.

For instance, free and quasi-free fields are described by Gaussian
states $F$. A generating function $Z$ of a Gaussian state $F$ on $B_\Phi$
reads
\begin{equation}
Z(\f)=\exp[-\frac12 M(\f,\f)] \label{032}
\end{equation}
where the covariance form $M(\f_1,\f_2)$ is a positive-definite
Hermitian separately continuous bilinear form on $\Phi$.
This generating function is the Fourier transform of a
Gaussian measure in $\Phi'$. The forms $F_{n>2}$ obey the Wick rules where
\[
F_1=0, \qquad F_2(\f_1,\f_2)=M(\f_1,\f_2).
\]
In particular, if $\Phi=\Bbb RS_n$, the covariance form of a Gaussian
state is uniquelly defined by a distribution $W\in S'(\Bbb R^{2n})$:
\[
M(\f_1,\f_2)=\int W(x_1,x_2)\f_1(x_1)\f_2(x_2)
d^nx_1d^nx_2.
\]

In field models, a generating function $Z$ plays the role of a
generating functional represented by the functional integral
(\ref{031}). If $Z$ is the Gaussian generating function (\ref{032}),
its covariance form $M$ defines Euclidean
propagators. Propagators of fields on the Mincowski space are
reconstructed by the Wick rotation of $M$ \cite{sard10}.

\section{Scalar fields}

In particular, set $\Phi=\Bbb RS_4$. Let $F$ be a Gaussian state on
the algebra $B_{\Bbb RS_4}$ such that its covariance form is represented
by a distribution $M(\f,\f')\in S'(\Bbb R^8)$ which is the Green function
of some positive elliptic differential operator
\[
L_yM(y,y')=\dl(y-y').
\]
This Gaussian state describes quasi-free Euclidean fields with the
propagator $M(y,y')$. For instance, if
\[
L_y =-\Delta_y+m^2,\qquad M(y,y')=\int\frac{\exp(-iq(y-y'))}{q^2+m^2}
d_4q,
\]
where $q^2$ is the Euclidean scalar, the Gaussian state $F$ describes
free Euclidean fields.

Note that, in case of the infinite-dimensional linear spaces, the measure
$\mu$ in the expression (\ref{031}) fails to be brought into explicit form.
The familiar expression
\begin{equation}
\exp(-S[\f]\op\prod_x [d\f(x)]\label{S11}
\end{equation}
is not a true measure. At the same time, the measure $\mu$ in $\Phi'$
is uniquely defined by the set of mesures $\mu_N$ in the finite-dimensional
spaces
\[
\Bbb R_N=\Phi'/E
\]
where $E\subset \Phi'$ denotes the subspace of forms on $\Phi$ which
are equal to zero at some finite-dimensional subspace $\Bbb
R^N\subset\Phi$. The measures $\mu_N$ are images of $\mu$ under the
canonical mapping $\Phi'\to\Bbb R_N$. For instance, every vacuum
expectation $F(\f_1\cdots\f_n)$ admits the representation by
functional integral
\begin{equation}
F(\f_1\cdots\f_n)=\int\langle w,\f_1\rangle\cdots\langle w,\f_n
\rangle d\mu_N(w) \label{S8}
\end{equation}
for any finite-dimensional subspace $\Bbb R^N$ which contains
$\f_1,\ldots,\f_n$. In particular, one can replace the generating
function (\ref{031}) by the generating function
\[
Z_N(\la_ie^i)=\int\exp(i\la_iw^i)\mu_N(w^i)
\]
on $\Bbb R^N$ where $\{e^i\}$ is a basis for $\Bbb R^N$ and $\{w^i\}$
are coordinates with respect to the dual basis for $\Bbb R_N$. If $F$ is
a Gaussian state, we have the familiar expression
\begin{equation}
d\mu_N=(2\pi\det[M^{ij}])^{-N/2}\exp[-\frac12(M^{-1})_{ij}w^iw^j] d^Nw
\label{S9}
\end{equation}
where $M^{ij}=M(e^i,e^j)$ is the non-degenerate covariance matrix.

The representation (\ref{S8}) however is not unique, and the measure
$\mu_N$ depends on the specification of the finite-dimensional subspace
$\Bbb R^N$ of $\Phi$.

Note that the Gaussian measure (\ref{S9}) admits the following
countably infinite generalization.

Let $\{e_i, i\in I\}$ be an orthonormal basis of $\Phi$ with respect to
some scalar form $\langle |\rangle$ on $\Phi$. Let $\Bbb R^I$ denotes a
linear space of real functions $\f$ on the set $I$ and $\Bbb R^I$ be
provided with the simple convergence topology. The collection of forms
\[
\e^i: \Bbb R^I\ni\f\to\f(i)\in\Bbb R, \qquad i\in I,
\]
is an algebraic basis of the dual $(\Bbb R^I)'$ isomorphic to the
subspace $\Phi_I\subset\Phi$ of vectors $\f$ meeting finite
decomposition
\[
\f=\la_ie^i
\]
with respect to the basis $\{e^i\}$ for $\Phi$. Then, the restriction
of a Gaussian generating function on $\Phi$ to $\Phi_I$ is the Fourier
transform of some Gaussian premeasure $\mu_I$ in $\Bbb R^I$ which is a
measure if the set $I$ is countable \cite{bur}. For example, if
\[
M(e^i,e^j)=\frac12\dl^{ij},
\]
the corresponding measure $\mu_I$ in $\Bbb R^I$ is
\begin{equation}
\op\prod_{i\in I} (\pi^{-1/2}\exp[-(w^i)^2]dw^i). \label{S12}
\end{equation}

The expression (\ref{S11}) in QFT is the formal continuum generalization of
the measure (\ref{S12}).

In contrast to the formal expression (\ref{8.1}) of perturbed QFT, the
true integral representation (\ref{031}) of generating functionals
enables us to handle non-Gaussian and nonequivalent Gaussian representations
of the algebra $B_\Phi$. Here, we describe one of such a representation
which is utilized to axiomatic theory of a Higgs vacuum \cite{S91}.

In contrast with the finite-dimensional case, Gaussian measures
in infinite-dimensional spaces  fail to be
quasi-invariant under translations as a rule. Let $\mu$ be a Gaussian
measure in the dual $\Phi'$ to a nuclear space $\Phi$ and $\mu_a$ the
image of $\mu$ under translation
\[
\g:\Phi'\ni w\to w+a\in\Phi', \qquad a\in\Phi'.
\]
The measures $\mu$ and $\mu_a$ are equivalent iff the vector
$a\in\Phi'$ belongs to the canonical image $\Phi^d$ of $\Phi$ in
$\Phi'$ with respect to the scalar form
\[
\langle |\rangle=M(,)
\]
where $M$ is the covariance form corresponding to the measure $\mu$.
In this case, we have
\[
\langle a,\Phi\rangle =\langle \Phi|\f_a\rangle
\]
for a certain vector $\f_a\in\Phi$. Then,
the measures $\mu$ and $\mu_a$ define equivalent
representations of the algebra $B_\Phi$. This equivalence is performed
by the unitary operator
\[
u(w)\to \exp(-\langle w|\f_a\rangle)u(w+a), \qquad u(w)\in
L^2(\Phi',\mu).
\]
This operator fails to be constructed if $a$ does not belong to $\Phi^d$.

\section{Fermion fields}

Chronological vacuum expectations of fermion fields are skew-symmetric.
By analogy with the scalar field case, we therefore can construct their
Euclidean images as forms on an anticommutative algebra.

Given a complex linear locally convex topological space $V$, let $C(V)$
be the quotient of the tensor algebra $A_V$ by the two-sided ideal with
the generating set
\[
J:=\{vv'+v'v,\qquad v,v'\in V\}.
\]
All finite-dimensional subspaces $\Bbb R^N$ of $V$ yields the inductive
family of subalgebras $\{C(\Bbb R^N)\}$ of $C(V)$. Every projective
family of continuous forms $F_N$ on $C(\Bbb R^N)$ defines some
projective limit form $F$ on $C(V)$, but this form fails to be
continuous in general.

By analogy with the scalar field case, one can consider the
homomorphism of the algebra $C(V)$ onto some algebra of functions in
order to construct $F$.

We denote by $\La$ a complex infinite-dimensional topologycal Grassmann
algebra
\[
\La=\La_0\oplus\La_1
\]
endowed also with the $\Bbb Z$-graded structure $\La=\prod\La^i$. The
algebra $C(V)$ has still the structure of a $\Bbb Z_2$-graded linear
space. Let us construct the $\La$ envelope $\La C(V)$ of $C(V)$. We
denote by
\[
\La V=\La_1\otimes V
\]
the $\La$ envelope of $V$ treated as the $\Bbb Z_2$-graded space
containing only an odd part. Both $\La V$ an $\La C(V)$ have structures
of $\La_0$-linear spaces. Let $\La\ol V$ be a $\La_0$-linear space of
$\La_0$-linear continuous mappings of $\La V$ into $\La$ which preserve
$\Bbb Z_2$-gradation. We further call the $\La$-valued form any linear
mapping into $\La$. The space $\La\ol V$ contains the $\La$ envelope of
the space $\ol V$ of $\La^1$-valued continuous forms on $V$. To give
analogy to the scalar field case, $\ol V$ is termed the $\La^1$-dual to
$V$ together with the pairing form $\langle;\rangle$.

One can endow $\ol V$ with the topology such that, whenever $v\in V$,
the form
\[
\ol V\ni s\to \langle s;v\rangle\in \La^1
\]
is continuous. Let us consider the correspondence between elements
$\{v^1\cdots v^n\}$ of $C(V)$ and continuous functions
\[
\langle s;v^1\rangle\cdots\langle s;v^n\rangle
\]
on $V$. This correspondence yields the homomorphisms
\[
c\to c(s), \qquad c\in C(V),
\]
of the algebras $C(V)$ and $\La C(V)$ into the algebras of continuous
$\La$-valued functions and $\La_0$-val;ued functions on $\ol V$
respectively.

Various $\La$-valued and numerical forms on different subalgebras of the
algebras of such functions can be constructed, e.g., by means of real
functions on $\La$ and measures in $\ol V$. We here examine forms whose
restrictions to $C(V)$ are numerical Gaussian forms. The corresponding
forms $F_N$ on $C(\Bbb R^N)$ can be represented by means of the Berezin
forms.

Given $\Bbb R^N\subset V$, there is homomorphism of $C(\Bbb R^N)$ onto
the subalgebra of $\La$-valued functions on
\[
\ol{\Bbb R}^N=\ol V/\ol E_N
\]
where the linear subspace $\ol E_N\subset\ol V$ consists of the forms
equal to zero on $\Bbb R^N\subset V$. Let $c_N$ be an annihilator element
of $C(\Bbb R^N)$ and $F_{BN}$ the Berezin form on $C(\Bbb R^N)$ given
by the expressions
\begin{equation}
F_{BN}(\{v^1\cdots v^{n<N}\}(s)=0, \qquad F_{BN}(c_N(s))=1. \label{F5}
\end{equation}
Remark that, for different annihilator elements, the forms $F_{BN}$ on
$C(\Bbb R^N)$ differs in numerical multipliers. For each element $z\in
C(\Bbb R^N)$, we then can define the form
\begin{equation}
F_{zN}(c)=F_{BN}(cz) \label{F^}
\end{equation}
on $C(\Bbb R^N)$. For instance, $F_{zN}$  is a nondegenerate Gaussian
form if
\[
z(s)=L\exp[S^{-1}_{ij}e^i(s)e^j(s)], \qquad L\in\Bbb R,
\]
where $S_{ij}$ is the matrix of a nondegenerate skew-symmetric bilinear
form on $\Bbb R^N$ (i.e., $N$ is even) with respect to a basis
$\{e^i\}$ for $\Bbb R^N$.

Note that, given the basis $\{e^i\}$, one can represent the form $F_{BN}$
by means of the Berezin-like integrals
\begin{equation}
F_{BN}=\int cd^Ne^i, \qquad c_N=e^1\cdots e^N. \label{F7}
\end{equation}
We here are not concerned of different approaches to superintegration.
Note only that the Berezin form (\ref{F5}) does not depend on
specification of a Grassmann algebra $\La$. The Berezin forms do not
preserve $\Bbb Z_2$-gradation. They fail to form a projective family
since, given $\Bbb R^n\subset \Bbb R^N$, we observe that
\[
F_{Bn}(c_n)\neq F_{BN}(jc_n), \qquad j: C(\Bbb R^n)\to C(\Bbb R^N).
\]
The expressions (\ref{F5}) - (\ref{F7}) which contain $F_{BN}$
therefore can not be extrapolated to the infinite dimension case. At
the same time, Gaussian forms $F_N$ can possess the continuous
projective limit $F$ on $C(V)$.

One can think of the algebra $C(V)$ as being the enveloping algebra of
the commutative Lie superalgebra corresponding to the Lie supergroup of
translations in the space $\La V$. This group can be represented by operators
\[
q(s)\to \exp[i\langle s;w\rangle]q(s), \qquad w\in \La V,
\]
in some subspace $Q$ ($\La C(V)\subset Q$) of $\La$-valued continuous
functions on $\ol V$. Given a continuous $\La$-form $F$ on $Q$ with
$F(1)=1$, this representation is characterized by the $\La$-valued
continuous function
\[
Z(w)=F(\exp[i\langle s;w\rangle]).
\]
Let the function
\[
\La^n\supset (\la_1,\ldots,\la_m)\to Z(\la_iv^i)\subset\La
\]
be holomorphic for every finite collection $v^1,\ldots,v^m\in V$.
Then, $Z(w)$ is a generating function for a form $F$ on $C(V)$ given by
the body parts of expressions
\[
F(v^1\cdots v^m)= i^{-m}\frac{\dr}{\dr\la_1}\cdots\frac{\dr}{\dr\la_m}
Z(\la_iv^i) =F(\langle s;v^1\rangle\cdots\langle s;v^m\rangle).
\]

For instance, if $V=\Bbb R^N$, a Gaussian form on $C(\Bbb R^N)$ is
defined by a generating function
\be
&& Z(\la_ie^i) = F_{BN}(L\exp[i\langle s;\la_ie^i\rangle +
S^{-1}_{ij}\langle s;e^i\rangle\langle s;e^j\rangle])=\\
&& \qquad \exp[-S^{ij}\la_i\la_j], \qquad L\in\Bbb R.
\ee

Let $V$ be a complex nuclear space endowed with an inner product $(,)$
and an involution operation such that
\[
(v^*,v'^*)=\ol{(v,v')}.
\]
Let $V$ be split into two orthogonal isomorphic subspaces
\[
V=H\oplus H^*.
\]
Every continuous bilinear skew-symmetric form $S(v,v')$ on $V$ is
equivalent to the form
\[
(v^*,S_-v')
\]
where $S$ is some bounded linear operator on $H$ and
\be
&&S_-v=Sv,\quad v\in H,\\
&&S_-v=-(S^+v^*)^*, \quad v\in H^*.
\ee
Note that $S$ is invertible if $S(v,v')$ is nondegenerate. A
skew-symmetric form $S(v,v')$ on $V$ yields a symmetric form $S(w,w')$
on $\La V$. A continuous Gaussian form $F$ on $C(V)$ is defined by a
generating $\La_0$-valued function
\begin{equation}
Z(\la_iv^i)=\exp[-\la_i\la_j(v^{i*},S_-v^j)].\label{F8}
\end{equation}

For instance, let $H$ be the space $\Omega\oplus S(\Bbb R^4)$ where by
$\Omega$ is meant the space of Euclidean 4-component spinors. Elements
of $\ol V$ then can be treated as superdistributions (which however
differ from that by ref \cite{nag}). If $S$ is the Green function of
the Euclidean Dirac operator, the generating function (\ref{F8})
describes free Euclidean fermion fields.

It should be noted that given a bilinear skew-symmetric form $S(v,v')$
on $V$, there exists the unique positive continuous form $M(v,v')$
\cite{kup} such that
\be
&& S(v,v')=M(Jv,v'),\\
&& M(v,v')=(x,A_1x')+(y^*,A_2y')
\ee
where
\begin{itemize}\begin{enumerate}
\item $v=x+y$ and $v'=x'+y'$ with $x,x'\in H$ and $y,y'\in H^*$;
\item $S=UA_1=A_2U$ is the polar decomposition of $S$ with a unitary
operator $U$ and positive operators $A_1$ and $A_2$;
\item $Jv=(U^+)_-v^*$ is an antiunitary operator with $J^2=-\id$.
\end{enumerate}\end{itemize}

For instance, if $S=\id$, we have
\[
M(v,v')=(v,v').
\]
Then, the Gaussian measure in $V'$ with the Fourier transform
\[
\exp[-\frac12M(v,v)]
\]
can be related with the generating function (\ref{F8}).

\section{Appendix}

\quad We use the Fourier-Laplace (F-L) transforms in order to construct
strictly the Wick rotation.

{\bf Remark}. By $\Bbb R^n_+$ and $\ol\Bbb R^n_+$, we
denote the subset of $\Bbb R^n$ with the Descartes coordinates
$x^\mu>0$ and its closure
respectively. Elements of $S(\Bbb R^n_+)$ are $\f\in S(\Bbb
R^n)$ such that $\f=0$ on $\Bbb R^n\setminus\Bbb R^n_+$. Elements
of $S(\ol\Bbb R^n_+)$ correspond to distributions $W\in S'(\Bbb R^n)$
with ${\rm supp}W\subset\ol\Bbb R^n_+$.

Given $W\in S'(\Bbb R^n)$, let $Q_W$ be the set of
$q\in\Bbb R_n$ such that
\[
\exp(-qx)W\in S'(\Bbb R^n).
\]
 The L-F transform of W is defined to be the Fourier transform
\[
W^{FL}(k+iq)=(\exp(-qx)W)^F=\int\exp[i(k+iq)x]Wd^nx\in S'(\Bbb R_n).
\]
It is a holomorphic function on the tubular set $\Bbb R_n+i{\op
Q^\circ}_W\subset\Bbb C_n$ over the interiority of $Q_W$. Moreover, it
defines the family of distributions $W^{FL}_q(k)\in S'(\Bbb R_n)$ which
is continuous in the parameter $q$. In particular, if $W\in S'(\ol\Bbb
R^n_+)$, then $\ol\Bbb R_{+n}\subset Q_W$ and $W^{FL}$ is a holomorphic
function on the tubular set $\Bbb R_n+i\Bbb R_{+n}$ so that
$W^{FL}(k+i0)=W^F(k)$, i.e.,
\[
\op\lim_{|q|\rightarrow 0, q\in\Bbb R_{+n}}\langle W^{FL}_q,\f\rangle=
\langle W^F,\f\rangle, \qquad \f(k)\in S(\Bbb R_n).
\]

Let $W^{FL}(k+iq)$ be the L-F transform of some distribution $W\in
S'(\ol\Bbb R^n_+)$. Then, the relation
\ben
&&\op\int_{\Bbb R_{+n}}W^{FL}(iq)\f(q)d_nq=
\op\int_{\ol\Bbb R^n_+}W(x)\wh\f(x)
d^nx, \qquad \f\in S(\Bbb R_{+n}), \nonumber\\
&&\wh\f(x)=\op\int_{\Bbb R_{+n}}\exp(-qx)\f(q)d_nq, \qquad
x\in\ol\Bbb R^n_+ \qquad \wh\f\in S(\ol\Bbb R^n_+), \label{035}
\een
defines the continuous linear functional $W^L(q)=W^{FL}(iq)$ on $S(\Bbb
R_{+n})$. It is called the Laplace transform. The image of $S(\Bbb
R_{+n})$ under the continuous morphism $\f\rightarrow\wh\f$ is dense in
$S(\ol\Bbb R^n_+)$, and the norms $\|\f\|'_{k,l}=\|\wh\f\|_{k,l}$
induce the weakening topology in $S(\Bbb R_{+n})$. The finctional
$W^L(q)$
(\ref{035}) is continuous with respect to this topology. There is the 1:1
correspondence between Laplace transforms of elements of
$S'(\ol\Bbb R^n_+)$ and elements of $S'(\Bbb R_{+n})$ continuous
with respect to the weakening topology in $S(\Bbb R_{+n})$. We use this
correspondence in order to construct the Wick rotation.

If the Minkowski space is identified with the real subspace $\Bbb R^4$ of
$\Bbb C^4$, its Euclidean partner is the subspace $(iz^0, x^{1,2,3})$
of $\Bbb C^4$. These spaces have the same spatial coordinate subspace
$(x^{1,2,3})$.
For the sake of simplicity, we further do not write spatial coordinate
dependence. We consider the complex plane $\Bbb C^1=X\oplus iZ$ of time
$x$ and Euclidean time $z$ and the complex plane $\Bbb C_1=K\oplus iQ$
of the associated momentum coordinates $k$ and $q$.

Let $W(q)\in S'(Q)$ be a distribution such that
\begin{equation}
W=W_++W_-,\qquad W_+\in S'(\ol Q_+), \qquad W_-\in S'(\ol Q_-).\label{036}
\end{equation}
For every $\phi_+\in S(X_+)$, we have
\ben
&&\op\int_{\ol Q_+}W(q)\wh\phi_+(q)dq=\op\int_{\ol
Q_+}dq\op\int_{X_+}dx [W(q)\exp(-qx)\phi_+(x)]=\nonumber\\
&&\qquad\op\int_{\ol Q_+}dq\op\int_Kdk\op
\int_{X_+}dx[W(q)\phi^F_+(k)\exp(-ikx-qx)]=\nonumber\\
&&\qquad -i\op\int_{\ol Q_+}dq\op\int_Kdk[W(q)\frac{\phi^F_+(k)}{k-iq}]=
\op\int_{\ol Q_+}W(q)\phi^L_+(iq)dq \label{037}
\een
due to the fact that the F-L transform
$\phi^{FL}_+(k+iq)$ of $\phi_+\in S(X_+)\subset S'(X_+)$
exists and it is holomorphic on the tubular set $K+iQ_+, Q_+\subset
Q_{\phi_+}$, so that $\phi^{FL}_+(k+i0)=\phi^F_+(k)$. The function
$\wh\phi_+(q)=\phi^{FL}_+(-q)$ can be regarded as the Wick rotation of
$\phi_+(x)$. The relations (\ref{037}) take the form
\ben
&&\op\int_{\ol Q_+}W(q)\wh\phi_+(q)dq=\op\int_{X_+}\wh W_+(x)\phi_+(x)dx,
\nonumber\\
&&\wh W_+(x)=\op\int_{\ol Q_+}\exp(-qx)W(q)dq, \qquad x\in X_+, \label{038}
 \een
where $\wh W_+(x)\in S'(X_+)$ is continuous with respect to
the weakening topology in $S(X_+)$.

For every $\phi_-\in S(X_-)$, we have the similar relations
\ben
&&\op\int_{\ol Q_-}W(q)\wh\phi_-(q)dq=
\op\int_{X_-}\wh W_-(x)\phi_-(x)dx,
\nonumber\\
&&\wh W_-(x)=\op\int_{\ol Q_-}\exp(-qx)W(q)dq, \qquad x\in X_-. \label{039}
\een

The combination of (\ref{038}) and (\ref{039}) results in the relation
\ben
&&\op\int_QW(q)\wh\phi(q)dq=\op\int_X\wh W(x)\phi(x)dx, \nonumber\\
&&\phi=\phi_++\phi_-, \qquad \wh\phi=\wh\phi_++\wh\phi_-,\label{040}
\een
where $\wh W(x)$ is a linear functional on functions $\phi\in S(X)$ such
that $\phi$ and all its derivatives are equal to zero at $x=0$. This
functional can be regarded as a functional on $S(X)$, but it
needs additional definition at $x=0$. This is the well-known feature of
chronological forms in quantum field theory. We can treat
$\wh W$ as the Wick rotation of $W$.

For instance, let the covariance form $\Lambda$ of a Gaussian state on
the commutative algebra of Euclidean
scalar fields $\wt\phi$ be given by a
distribution $\wt W(z_1-z_2) $. We have
\be
&&\int\wt W(z_1-z_2)\wt\phi_1(z_1)\wt\phi_2(z_2)
dz_1dz_2=
\int\wt w(z)\wt\phi_1(z_1)\wt\phi_2(z_1-z)dz_1dz=\\
&&\qquad \int\wt w(z)\wt f(z)dz=\int\wt w^F(q)\wt f^F(q)dq,\\
&&z=z_1-z_2, \qquad \wt f=\int\wt\phi_1(z_1)\wt\phi_2(z_1-z)dz_1.
\ee
Let $\wt w^F(q)$ satisfy the condition (\ref{036}). Its Wick rotation
(\ref{040}) defines the functional
\be
&&\wh w(x)=\theta(x)\op\int_{\ol Q_+}\wt w^F(q)\exp(-qx)dq+
\theta(-x)\op\int_{\ol Q_-}\wt w^F(q)\exp(-qx)dq\\
&&\int\wh w(x)f(x)dx=\int\wh W(x_1-x_2)\phi(x_1)\phi(x_2)dx_1dx_2,
\ee
on scalar fields $\phi$ on the Minkowski space.
For instance, if $\wt w^F(q)$ is
the Feynman propagator, $i\wh W$
is the familiar causal Green function $D^c$.

\end{document}